\documentclass[fleqn,10pt]{wlscirep}
\usepackage[utf8]{inputenc}
\usepackage[T1]{fontenc}
\usepackage{breqn}
\newcommand\bib{\bibitem}

\title{ {\bf A Quantum Field Theoretical Study of Correlated Quantum Ising model with 
		Longer Range Interaction}}
\author[1,2,$\dagger$]{Ranjith R Kumar}
\author[1]{Sujit Sarkar}
\affil[1]{Department of Theoretical Sciences, Poornaprajna Institute of
	Scientific Research, 4, Sadashivanagar, Bangalore-560 080 and Bidalur Post
	, Devanhalli, Bangalore Rural 562110, India.}
\affil[2]{Graduate Studies, Manipal Academy of Higher Education, 
	Madhava Nagar, Manipal-576104, India.}
\affil[$\dagger$]{ranjith.btd6@gmail.com}
\affil[*]{sujit.tifr@gmail.com}

\begin{document}
\begin{abstract}
	The physics of quantum Ising model (qIm) plays an important role in quantum 
	many body system. 
	We study and present the results of qIm and longer range quantum 
Ising model (lqIm) 
in presence of strong correlation.
	We do the quantum field theoretical renormalization
	group (RG) calculation to study the behaviour of RG
	flow lines for different couplings for different region of parameter space.
	We show how the strong correlation
	effect enrich the quantum physics of these two systems.
	We show explicitly that the ordered ferromagnetic (FM) phase to 
	the disorder quantum paramagnet (dqpI) quantum phase transition 
	occurs for only in the strongly correlated regime for qIm and the
	dqpI phase appears for non-interacting and attractive regime.
	We show explicitly for lqIm that FM to dqpI transition
	occurs at the extremely correlated region and also the dqpI phase appears 
in correlated regime. 
	We show that 
	short range FM coupling and 
	longer range coupling 
are competiting with each other and also the effect of strong correlation in this competition.
	We also show the most interesting feature that the transverse field oppose the
	FM coupling of qIm but it is favour the longer range coupling of lqIm. We find the
evidence of another disorder quantum paramagnetic (dqpII) phase due to the relevance
of longer range coupling. We also present the existence of another quantum phase 
transition from dqpII phase to FM phase. We show explicitly that there is no 
phase transition 
from dqpI phase to dqpII phase rather they coexists.
	This work provides a new perspective not only for the statistical physics of 
quantum Ising model but also
	for the quantum many body systems.  
\end{abstract}
\maketitle

\section{Introduction}
\noindent 
The physics of quantum Ising model (qIm) plays an important role in one dimensional quantum 
many body system $^{1-8}$. The physics of one dimensional quantum many body is interesting in its
own right $^{9,10}$. 
One dimensional quantum many body system has strong quantum fluctuations that do not
allow spontaneously broken continuous symmetries 
As a result of this, the pairing instabilities do not lead to 
any 
ordered density-wave $^{11,12}$. This many body has a critical phase with power law decay of various
correlation functions, universally known as a Luttinger liquid. In one dimensional quantum
many body systems whether it is weakly correlated or strongly proper treatment of the
quantum fluctuations leads in both cases to a Luttinger liquid characterized by phonon-like
collective density fluctuation modes. In this process there is a one-to-one correspondence
between excitations of a one dimensional free Fermi gas and free bosons. This can be used
to solve the strongly interacting fermion problem by turning it into a weakly interacting
boson problem via the method of bosonization followed by the quantum field theoretical
renormalization group study $^{13-16}$. In the present study, we use the bosonization process to
recast the model Hamiltonian in continuum field theory followed by the renormalization
group (RG) method.\\
It is well
known that the Coulomb interaction is always present in the solid state, either
screened or weak or sometimes even stronger. Coulomb interaction leads to
the different physical phenomena in quantum many-body systems, such as the Kondo
effect, Mott-Hubbard transition and superconductivity to mention a few $^{9,10}$. Therefore,
to get a complete picture of emergent quantum phases of a quantum many-body system,
one has to consider the
effect of correlation $^{9,10}$.
In the correlated many-body system, this situation is described by the celebrated
sine-Gordon model $^{9}$, which also plays a central role in quantum field
theory.
In the low energy limit, effective degrees of freedom give the accurate description 
of the system . It is one of the method  by
which one find the low energy effective theory.\\ 
The mathematical structure and results of the RG theory
are a significant conceptual advancement in quantum field theory in
the last several
decades in both high-energy and quantum many body condensed matter physics.
The need for RG is
really transparent in condensed matter physics.
RG theory is a formalism
that relates the physics at different length scales in condensed matter physics
and the physics at different energy scales in high-energy physics $^{17,18}$.
We focus on systems that can be mapped to a dual-field double sine-Gordon model as a bosonized
effective field theory.
In this study we do the quantum field theoretical 
RG calculation for the correlated quantum Ising model (qIm) and
longer range quantum Ising model (lqIm).
We want to study this problem from the perspective of one dimensional correlated 
quantum many body system.\\
{\bf Motivation of this study} \\  
The physics of qIm has already studied extensively in the literature $^{1-7}$. 
But still the physics
of strong correlation has not explored for qIm. The quantum field theoretical study of our 
model Hamiltonian has not explored in the literature, specially how the strong correlation
physics explore for this model Hamiltonian. In the literature quite a few studies have
already been done for this model Hamiltonian from the perspective of topological
state. But the main motivation of our study is to explore
the physics from  
the perspective of correlated quantum many body system.      
This model Hamiltonian has three competiting
interaction terms, the most interesting feature of this study is 
to show how the behaviour of RG flow lines reflect for   
these three competiting interaction and finally leads to emergence phase.\\
\section{Model Hamiltonian and Renormalization Group Equation:}
The model Hamiltonian$^{19}$ of the present study is
\begin{equation}
H = - \sum_{i} (\mu  {{\sigma}_i }^{x} 
+ {\lambda}_1 {{\sigma}_{i} }^z {{\sigma}_{i+1}}^{z} 
+ {\lambda_2 }
{{\sigma}_i }^{x} {\sigma_{i-1}}^z {\sigma_{i+1}}^{z} ) .
\end{equation}
Here $\mu $ is the chemical potential, $\lambda_1 $ is the two spin interaction
of nearest-neighbour (NN) sites and $\lambda_2$ is the three spin interactions. 
Thus we term this model Hamiltonian as 
lqIm with    
three spin interaction.\\ 
It is well known to us for qIm, the system is in disorder quantum phase when
the transverse field exceed the FM coupling, we term this disorder quantum phase
as dqpI.  
The coupling $\lambda_2 $ is related with the three sites (i-1, i and i+1 ), the
left (i-1) and right (i+1 ) sites are related with the $\sigma_z $ operator and 
the middle site (i) is with the $\sigma_x $ operator $\sigma_x $ operator. It is
very clear from this interaction term that next-nearest-neighbour (NNN) sites are related
with the FM interaction but NN sites are related with the $XZ$ interaction, i.e, 
the spin flipping occurs at the site i. This interaction introduce the frustration
in the system and finally leads to the disorder quantum phase. We term this disorder
quantum phase as dqpII. We will see after the derivation of quantum field theoretical 
RG equations that the behaviour of the RG flow lines for the couplings $\mu$ and $\lambda_2 $
are the same and this result is also supported from the study of scaling relation.
Thus in this quantum many body system possesses two different kind of disorder quantum
phases. \\             
The whole Hamiltonian (eq.1) have been also  
        studied previously in different contexts 
        The model was first introduced by the authors of Ref. 19 
        to study the persistence of quantum criticality at high temperature 
in correlated systems. 
        The authors of Ref.20 have studied the physics 
        of Majorana zero modes in the gapped phases of this model with both broken and 
        unbroken time-reversal symmetry. 
        One of the authors (S.S) has studied the quantization of geometric 
phase  with 
integer and fractional 
        topological characterization for this model in Ref.21. 
        Very recently authors of Ref.22 have 
solved the problem of bulk-boundary correspondence
        at the quantum critical lines and discussed the principle of least topological 
invariant 
        number at the criticality. The author Ref. 23 have also studied
Curvature RG of this model Hamiltonian. But the quantum field theoretical RG
study and the interpretation of emergent quantum phases from the perspective of quantum
many body physics is absent in the literature.\\

The model Hamiltonian can be described by the
sine-Gordon field theory as,
\begin{equation}
H = H_{0} + V( \phi, \theta ),
\end{equation}
where  $H_{0} $ is
\begin{equation}
H_0 =
 (\frac{{h} v}{2 \pi}) \int dx [
{({\partial_x  \theta (x)})}^2 +
{({\partial_x  \phi (x)})}^2 ].
\end{equation}
The Hamiltonian, $H_0 $ gives a
universal framework for describing one dimensional interacting bosons and fermionic
system, i.e.,
Tomonoga-Luttinger liquid (TLL) Hamiltonian and $V (\phi)$ is the sine-Gordon potential.
$\theta (x)$ is the dual field of $\phi (x)$  and satisfy the following
commutation relation
, $ [ \phi (x), \partial_x \theta (x') ] = -i \pi \delta (x - x')$.
$v$ is the velocity of the collective excitation of the system.
We write final form of Bosonized Hamiltonian as (detail derivation is relegated to
to the "Method" section),
\begin{multline}
V(\phi , \theta)= \frac{\lambda_1}{2}\int \cos[4\sqrt{\pi K}\phi(x)]dx - \mu \int 
\cos[2\sqrt{\pi K }\phi(x)] \cos[\sqrt{\frac{\pi}{ K}}\theta(x)] dx\\ 
-\frac{\lambda_2}{\pi} 
\int \cos[2\sqrt{\pi K}\phi(x)] \cos[\sqrt{\frac{\pi}{K}}\theta(x)] 
\left( \partial_x \sqrt{K} \phi(x) 
\right)^2 dx 
\end{multline}
The bosonized form of the model Hamiltonian consists four terms. The first term
is the kinetic energy term and the rest three terms present the sine-Gordon 
coupling terms. It is to be noted that the starting Hamiltonian (eq. 1) has no
$K$, term but it appears after the continuum
field theoretical calculation in the bosonized version of the Hamiltonian
(detail derivation is relegated to the "Method" section). 
$K$ is
the Tomonoga-Luttinger liquid (TLL) parameter to present the interaction
strength in the system.
The physics of low-dimensional quantum many body condensed matter system
is enriched
with its new and interesting emergent behavior.
$K < 1$ and $K > 1$ and $ K=1 $ characterizes
the repulsive, attractive interactions and non-interacting, respectively $^{9,24,25}$.\\
We notice that the nearest-neighbour (NN) coupling term ($\lambda_1 $) 
is related with
the a single sine-Gordon coupling term of the field ($\phi $ (x) ).  
The transverse field is
the product of two sine-Gordon coupling terms of $\phi (x)$ and $\theta (x) $ 
which are dual
to the each other. But we notice that for the longer range coupling with three spin 
interaction is 
also product of two
sine-Gordon coupling terms augmented with a part of the kinetic energy term from the 
field $\phi (x)$. These extra two sine-Gordon coupling terms for the 
longer range interaction 
over the quantum Ising model give the enrich quantum physics over the qIm.\\
It is very clear from the continuum field theoretical study that our model 
Hamiltonian contains three
strongly relevant and mutually nonlocal perturbations over the 
Gaussian (critical) theory. In such a situation,
the strong coupling fixed point is usually determined by the most 
relevant perturbation whose amplitude grows
up according to its Gaussian scaling dimensions and it is not much affected 
by the less relevant coupling terms.
However, this is not the general rule if the operators exclude each other. 
In this case, the interplay between the
three competing relevant operators (here $\mu$, $\lambda_1 $ and 
$\lambda_2 $) are the three competing 
relevant operators, which are related with
dual fields $\theta (x) $ and $\phi (x) $ can produce a novel quantum phase 
transition through 
a critical point or a critical line $^{24,25}$.
Therefore, the present study based on RG equations will give us the appropriate 
results for these model Hamiltonian. \\
\begin{figure}
\includegraphics[scale=0.3,angle=0]{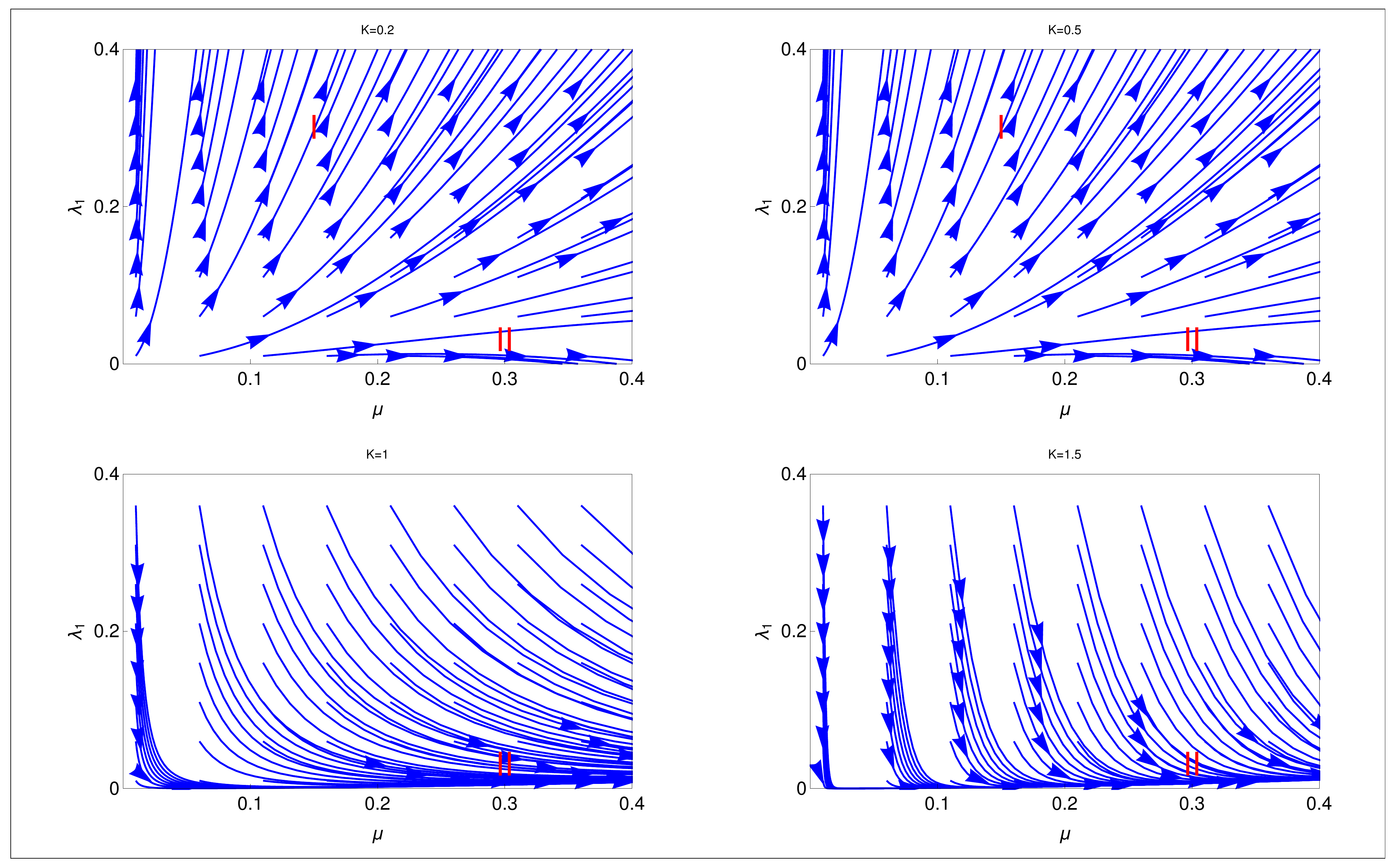}
\caption{ (Color online.)
This figure shows the behaviour of renormalization group flow lines for the couplings 
$\lambda_1 $ and $\mu$ (eq. 5) for the different initial values of $K$ as depicted in
figures.
}
\label{Fig. 1 }
\end{figure}
Now we present the RG equations for the present study:\\
(1). The RG equation for the correlated qIm is the following 
(detail derivation is related to the "Method" section ),
\begin{align}
\frac{d\lambda_1}{dl} &=
(2-4K )\lambda_1+ \frac{\mu^2}{8} (2K-\frac{1}{2K} ) \nonumber \\  
\frac{d\mu}{dl} &= (2-K-\frac{1}{4K} )\mu +\lambda_1\mu K \nonumber \\  
\frac{dK}{dl} &=  -\lambda_1^2 K^2
\end{align}
(2). The RG equation for the correlated lqIm is the following,
(detail derivation is related to the "Method" section ),
\begin{align}
\frac{d\lambda_1}{dl} &=
(2-4K )\lambda_1+ \frac{\mu^2}{8} (2K-\frac{1}{2K} ) \nonumber \\  
\frac{d\lambda_2}{dl} &=
(2-K-\frac{1}{4K} )\lambda_2 + \frac{\lambda_1\lambda_2 K}{\pi} \nonumber \\
\frac{d\mu}{dl} &= (2-K-\frac{1}{4K} )\mu +\lambda_1\mu K \nonumber \\  
\frac{dK}{dl} &= -\lambda_1^2 K^2 
\end{align}
(3). RG equation for the correlated lqIm without transverse field ($ \mu =0 $). \\
(detail derivation is related to the "Method" section ),
\begin{align}
\frac{d\lambda_1}{dl} &=
(2-4K )\lambda_1  \nonumber \\  
\frac{d\lambda_2}{dl} &=
(2-K-\frac{1}{4K} )\lambda_2 + \frac{\lambda_1\lambda_2 K}{\pi} \nonumber \\
\frac{dK}{dl} &= -\lambda_1^2 K^2 
\end{align}
The most interesting feature of these three set of RG equations is the 
appearance of K, therefore the behaviour of the RG
flow lines will be affected by the strong correlation. \\ 
\begin{figure}
	\includegraphics[scale=0.3,angle=0]{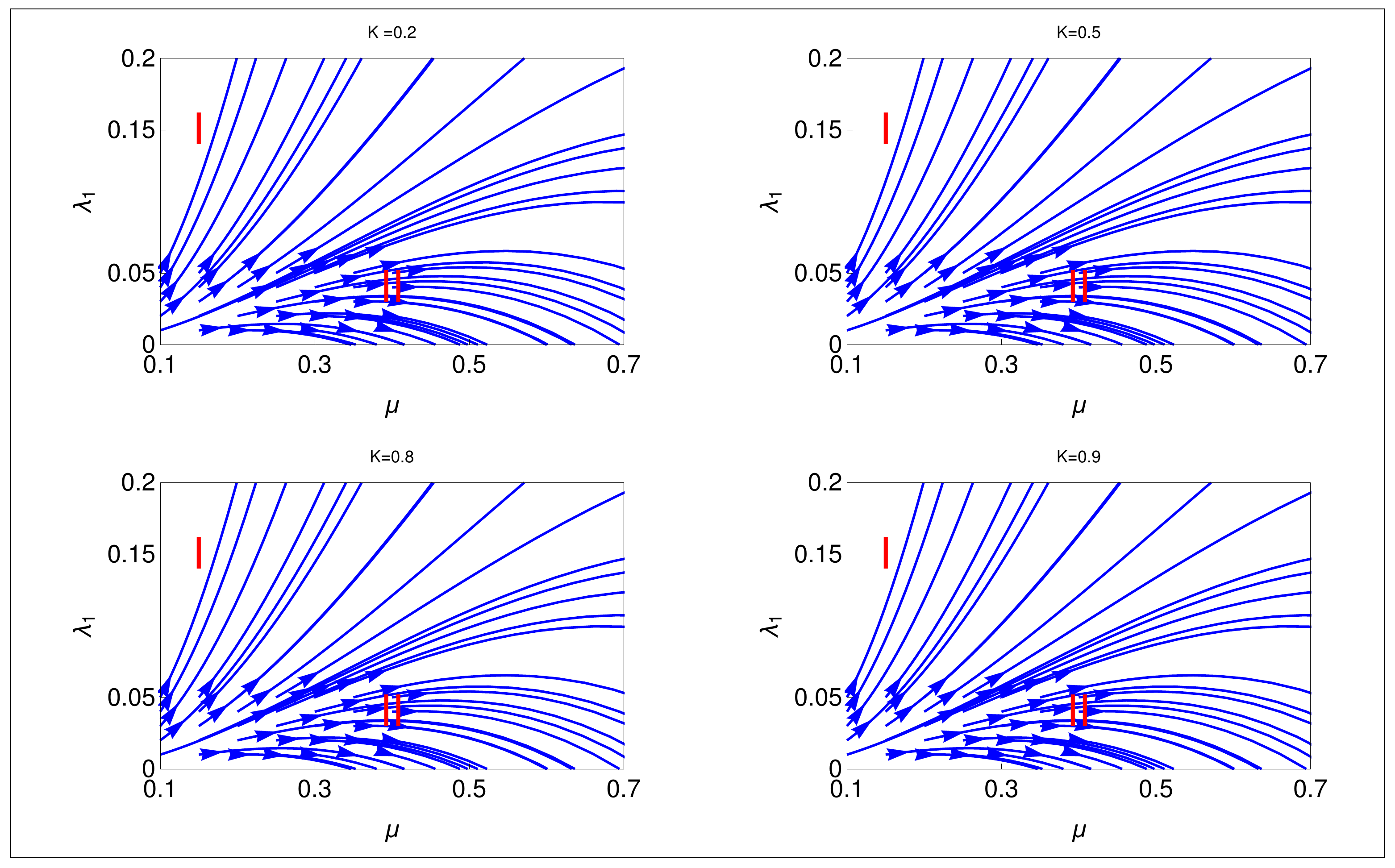}
	\caption{ (Color online.)
		This figure shows the behaviour of renormalization group flow 
lines for 
the couplings 
		$\lambda_1 $ and $\mu$ (eq. 5) for the different initial values of $K$ as depicted in
		figures.
	}
	\label{Fig. 2 }
\end{figure}
\section{Results:}
{\bf (A). Effect of strong correlation in quantum Ising model } \\
Fig. 1, presents the RG flow diagrams for the couplings $\lambda_1 $ and $\mu$
from the study of RG flow equation (eq. 5). 
It 
consists of four figures for different values of $K$ as depicted in the figures.
The figures for the first panel are for strongly correlated regime
, i.e., the $K <1$. The most interesting feature that we observe from these behavior
of RG flow lines is that there is a transition from the ordered FM phase to
dqpI phase. 
The behaviour of RG flow lines are the same for the entire correlated regime.
It reveals from this study that both the coupling $\lambda_1$ and $\mu$ are flowing
off to the strong coupling phase but the coupling $\lambda_1 $ 
increases more sharply than the smaller initial values of $\mu$.
But it reveals from the behaviour of RG flow lines
for smaller initial values of $\lambda_1 $ the RG flow lines are flowing off to the
weak coupling phase finally touches the base line, i.e., the system is in the
dqpI phase, we term this phase as dqpI. 
We find the quantum phase transition from the
ordered FM phase to dqpI phase to search this quantum phase transition in the
repulsive regime of strongly correlated phase. Thus we bench mark the standard
results of qIm of literature $^{1-7}$.\\ 
In fig.2 we show it explicitly from perspective of correlated physics. \\ 
The lower panel presents the results for non-interacting
($K=1$) and attractive regime ($ K> 1 $). The behaviour of the RG flow lines for
the coupling, $\lambda_1 $ is flowing off to the weak coupling phase and finally 
touch the base line.  
For this situation system is in the 
dqpI phase.
For these two figures, we have not found any quantum phase transition between the order 
FM phase to dqp phase. 
The RG flow lines are 
much more stiffer for the attractive regime of the parameter space.\\ 
\begin{figure}
\includegraphics[scale=0.6,angle=0]{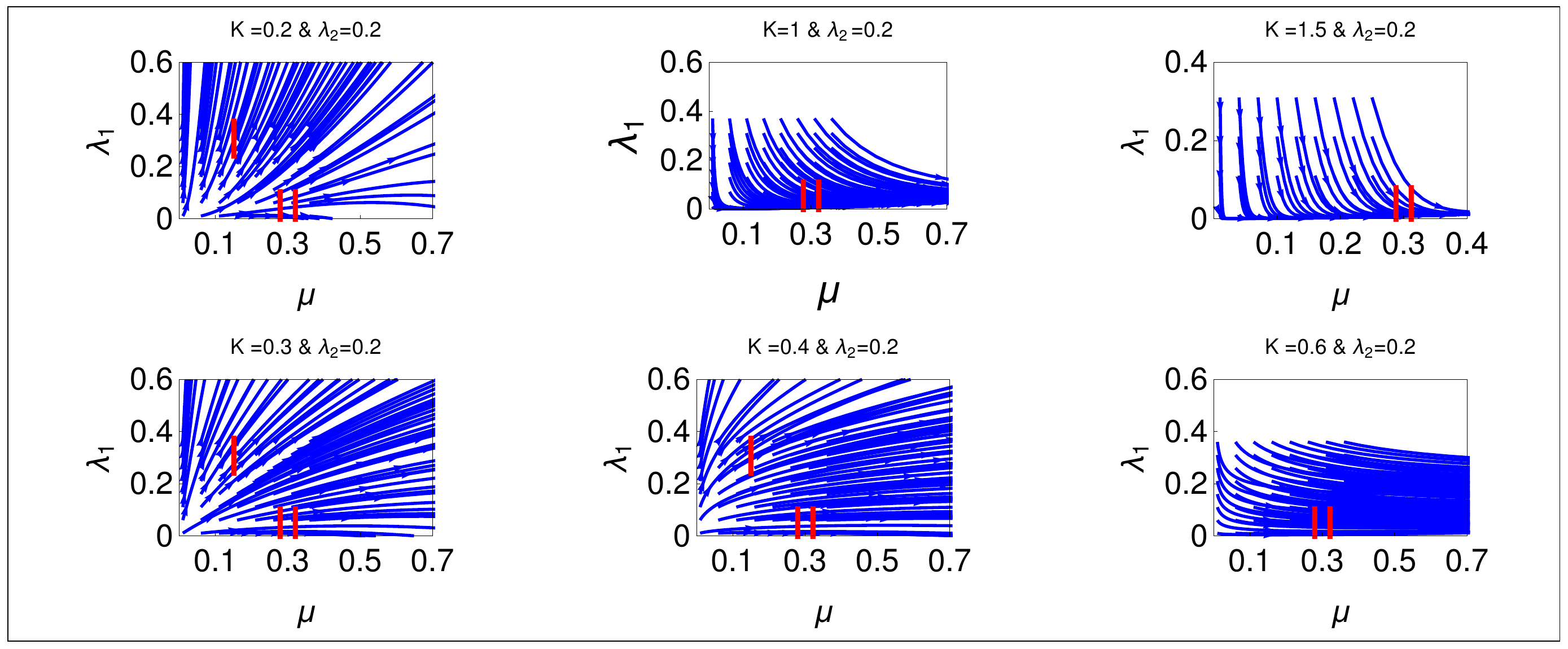}
\caption{ (Color online.)
This figure shows the behaviour of renormalization group flow lines for the couplings 
$\lambda_1 $ and $\mu$ (eq. 6) for the different initial values of $K$ and $\lambda_2 $ 
as depicted in
the figure.
}
\label{Fig. 3 }
\end{figure}
\begin{figure}
\includegraphics[scale=0.6,angle=0]{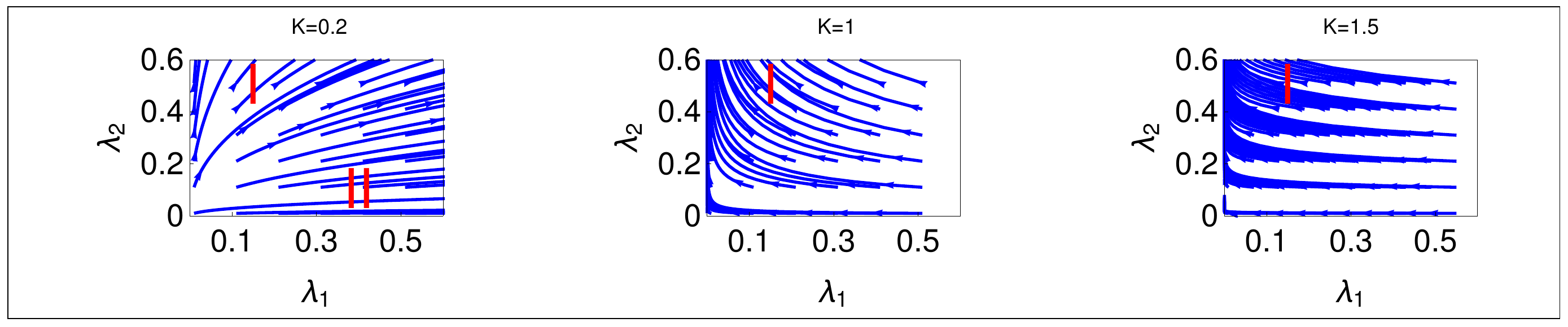}
\caption{ (Color online.)
This figure show the behaviour of renormalization group flow lines for the couplings 
$\lambda_2 $ and $\lambda_1 $ (eq. 7) for the different initial values of $K$
as depicted in figures.
}
\label{Fig. 4 }
\end{figure}
Fig. 2, presents the RG flow diagrams for the couplings $\lambda_1 $ and $\mu$
from the study of qIm (eq. 5). 
It 
consists of four figures for different values of $K$ as depicted in the figures.
The all figures are for strongly correlated regime
, i.e., the $K <1$. The main theme   
of this figure is to show the evidence of order FM to dqp phase transition explicitly.
The region (I) is the order FM phase and region (II) is the dqpI phase, 
the behaviour of
the RG flow lines are the same for the all initial values of $K$ in the correlated
regime. One of the hall-mark of this study is that only  
quantum phase transition occurs in the strongly correlated regime ,i.e., there is
no evidence quantum phase transition for the non-interacting and attractive regime.
Thus it is clear from this study that for the higher values of $K$, system
prefer to stay in the dqpI phase. 
Now the prime task
is to find the effect of longer range interaction on the qIm RG flow study.\\

{\bf (B). Effect of strong correlation  
in longer range quantum Ising model} \\
In fig.3, we present the RG flow lines behaviour from the study of 
coupling $\lambda_1 $ and $\mu$
for the RG equation (eq. 7). This figure consists of two panels, for all panels 
one we fix
the initial value of 
,$\lambda_2 =0.2 $, but for the different initial values of $K$. 
In the upper panel, there are three figures. The left,
middle and right figures are respectively for the initial values of $K = 0.2$,
$1$ and $1.5$. 
But in the lower panel  
we vary the initial value of $K$ for only the correlated regime
as depicted in the figures, 
to search the effect of $\lambda_2 $ on the quantum phase transition.
The most interesting result we obtain from the behaviour of RG flow lines for
the left figures of upper and lower panel that the quantum phase 
transition occurs at the extremely correlated regions, i.e., for the very lower
values of $K$ otherwise system is in the dqpI phase. It reveals from the
behaviour of RG flow lines for the middle and right figure of the upper panel that 
the RG flow lines flowing off to the weak coupling phase much sharper for the higher
initial values of $K$. \\ 
In the lower panel, the behaviour of RG flow lines for the
right figures show that the RG flow lines are flowing off to the 
weak coupling phase very slowly but not touching 
the base line. Here there is no evidence of quantum phase 
transition, system is only in the  order FM phase due to the relevant of coupling
$\lambda_1 $. The middle figure shows that order FM phase in region I and the
existence of flat phase, i.e., the RG flow lines move with the same initial
velocity in region II. \\
The most interesting result which we have obtained from this study that the
presence of $\lambda_2 $ drives the dqpI phase at the strongly correlated
regime. 
The competition between the FM coupling of qIm and long range coupling 
favour the
transverse field and drives the system to the dqpI phase at the extremely correlated regime. 
But for the left and middle 
figure we are unable to find the sharp quantum phase transition from order FM phase to
dqpI phase. For the smaller values of $\lambda_1$, the behaviour of RG flow lines
are almost flat for the middle figure. \\ 
\begin{figure}
\includegraphics[scale=0.4,angle=0]{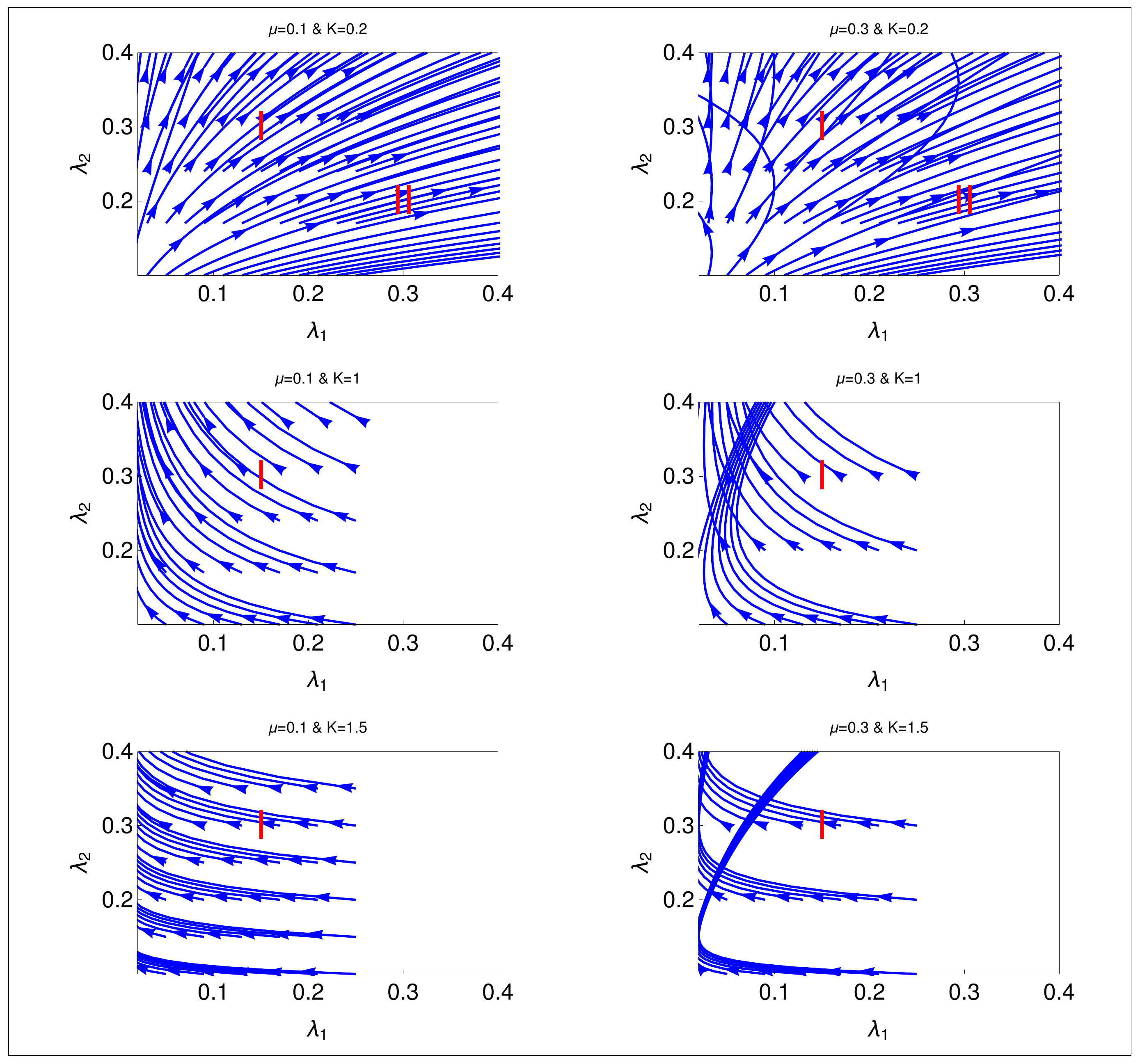}
\caption{ (Color online.)
Shows the behaviour of renormalization group flow lines for the couplings 
$\lambda_2 $ and $\lambda_1$ for the different initial 
values of $K$ and $\mu$.
This figure consists of three rows for different initial values of $K $.     
We present the RG flow lines based on eq.6. The left and right figures are
respectively for $\mu =0.1$ and $\mu =0.3$.    
}
\label{Fig. 5 }
\end{figure}
\begin{figure}
	\includegraphics[scale=0.6,angle=0]{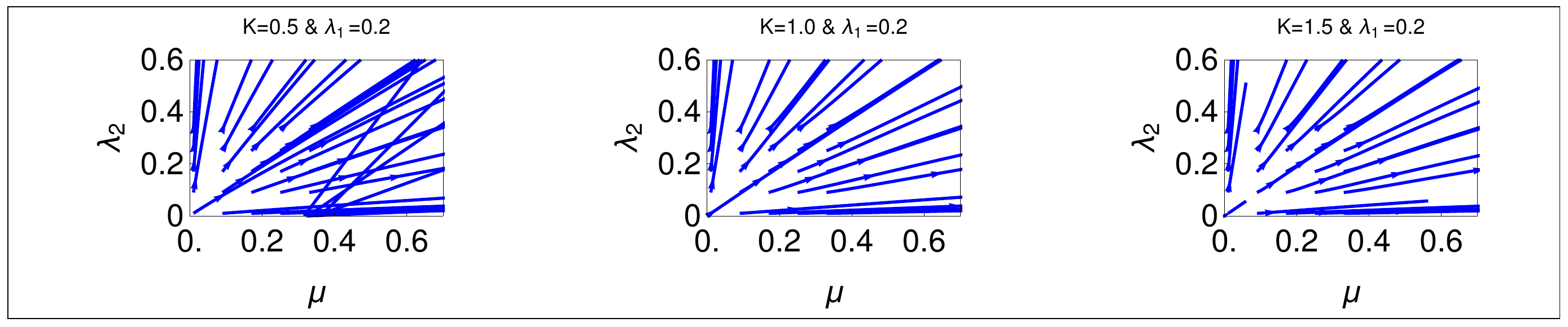}
	\caption{ (Color online.)
		This figure shows the behaviour of renormalization group flow lines for the 
couplings 
		$\lambda_2 $ and $\mu$ (eq. 7) for the different initial values of $K$ and $\lambda_1 $
		as depicted in figures.
	}
	\label{Fig. 6 }
\end{figure}
\begin{figure}
	\includegraphics[scale=0.5,angle=0]{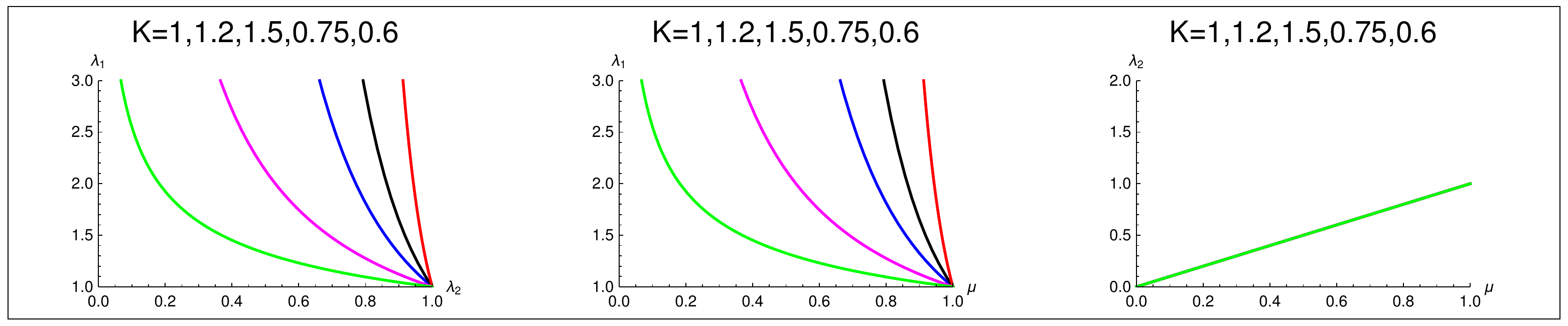}
	\caption{ (Color online.)
		Shows the results for scaling analysis  
		based on the equations (8) and (11). The Green, Magenta, 
Blue, Black and Red lines
		are respectively for $ K= 0.6, 0.75, 1, 1.2$ and $1.5$.  
	}
	\label{Fig. 7 }
\end{figure}
Fig. 4 presents the RG flow diagrams for the couplings $\lambda_2 $ and $\lambda_1 $ 
from the study of qIm (eq. 7) in absence of transverse field. 
It 
consists of three figures for different values of $K$ as depicted in the figures.
The left, middle and right figures are respectively for the strongly correlated,
non-interacting and attractive regime. 
It give us only how the $\lambda_1$ and $\lambda_2 $  
compete with each other in absence of transverse field. 
It reveals from the behaviour of RG flow lines that for the correlated regime both of 
the couplings are flowing off to the strong coupling phase, but when the initial values
of $\lambda_1 $ is greater than $\lambda_2 $, the RG flow lines for $\lambda_1 $ flowing off
more sharply than $\lambda_2 $, i.e., the system is in the 
ordered FM phase for $\lambda_1$ coupling otherwise the system is in disorder quantum phase due to
$\lambda_2 $ coupling. 
The middle and right figures show that the coupling, $\lambda_2 $, flowing off to the strong 
coupling phase and the system is in the disorder quantum phase due to the $\lambda_2 $ coupling. 
The origin of this disorder quantum phase is 
entirely different from the dqpI phase from the qIm, we term this
dqp phase as dqpII.\\
Fig. 5 presents the RG flow diagrams for the couplings $\lambda_2 $ and $\lambda_1 $ 
from the study of lqIm (eq. 6). Here we study the effect of transverse field on the
RG flow lines of fig.4. This figure consists of three panels the upper, middle and lower
panels are respectively for correlated ($K=0.2$), non-interacting ($K=1 $)  
and attractive regime ($K =1.5$). Each panel consists of two figures for two
different values of $\mu$, the left and right figures for each panel are respectively
for $\mu =0.1$ and $0.3$. \\
It reveals from the RG flow lines that for higher initial values of $\lambda_2$, i.e,
when the initial values of $\lambda_2 > \lambda_1 > $, 
the RG flow lines for the coupling $\lambda_2 $
increase more sharply than the coupling $\lambda_1 $ then the system is in 
dqpII phase. The situation is different when the initial values of 
$ \lambda_1  > \lambda_2 $,
for this case system is in the FM phase due to the $\lambda_1$ coupling, 
qualitative behaviour 
of the RG flow lines are the same. In the middle and lower panel, we observe that    
the RG flow lines are flowing off to the strong coupling phase due to the 
coupling $\lambda_2 $,
it increases very sharply for the higher values of $\mu$. 
Thus we finally conclude that the
presence of $\mu$ help the coupling $\lambda_2$ flowing off to the strong coupling phase, i.e.,
the system is in strong coupling phase.\\ 
Fig. 6 shows the RG flow diagrams for the couplings $\lambda_2 $ and $\mu $ 
from the study of qIm (eq. 7). 
This figure panel  
consists of three figures for different values of $K$ as depicted in the figures, here we
fix $\lambda_1 =0.2 $. The main theme of this study is to show whether there is any competition
between the $\mu$ and $\lambda_2 $. The analytical expressions for
the RG equations for these two couplings are the same at the one loop level   
but it differs in   
the second loop level. It reveals from the behaviour of RG flow lines that both
of the couplings are flowing off to the strong coupling phase almost at the same speed.
Thus it is clear that there is no quantum
phase transition and both dqpI and dqpII phases coexist. We have already explained 
that the nature of coupling $\lambda_2 $ is disorder quantum frustrated due to the 
three sites structure of the interaction.  \\
We observe from these study that dqpI phase exist only for the qIm and lqIm model 
when we find the competition between the order FM to disorder quantum paramagnetic 
phase. The dqpII phase exists only for lqIm when we study the RG flow lines for 
the couplings $\lambda_2 $ with $\lambda_1 $. The effect of strong correlation is
the same for the both qIm and lqIm, as we notice that the system prefers to stay
in dqpI phase for the higher values of $K$. \\    
\section{Scaling Analysis:}
It is well known that the critical theory is invariant under
the rescaling. Then the singular part of the free energy density satisfies the
following scaling relations $^{1}$.\\
\begin{align}
f_s [\mu, \lambda_1 , \lambda_2] = {\lambda_1 }^{2/(2 - 4K)}
f_s [1 ,
{\lambda_1 }^{-(2-4 K)/(2-K - 1/4K )} \lambda_2 ,
{\lambda_1 }^{-(2-4 K)/(2-K - 1/4K )} \mu
]  
\end{align}

The scaling equation for $\lambda_1 $ and $\lambda_2 $ is the following:\\ 
\begin{align}
f_s [\lambda_1 , \lambda_2] = {\lambda_1 }^{2/(2 - 4K)}
f_s [1 ,
{\lambda_1 }^{-(2-4 K)/(2-K - 1/4K )} \lambda_2
]  
\end{align}

The scaling equation for $\lambda_1 $ and $\mu $ is the following:\\ 
\begin{align}
f_s [\lambda_1 , \mu] = {\lambda_1 }^{2/(2 - 4K)}
f_s [1 ,
{\lambda_1 }^{-(2-4 K)/(2-K - 1/4K )} \mu
]  
\end{align}

The scaling equation for $\lambda_2 $ and $\mu $ is the following:\\ 
\begin{align}
f_s [\lambda_2 , \mu] = {\lambda_2 }^{2/(2 -K -1/4K)}
f_s [1 ,
{\lambda_2 }^{-(2- K- 1/4 K)/(2-K - 1/4K )} \mu
]  
\end{align}
In fig.7, we present the results of scaling relation based on the eq.8 to 11. This
figure panel consists of three figures. The left, middle and right figures are 
respectively 
for the scaling of $\lambda_1 $ with $\lambda_2 $, $\lambda_1 $ with $\mu$ and
$\lambda_2 $ with $\mu$. Each figure consists of five curves for different values of $K$ as
mention in the figure caption.\\
In the left figure for the correlated regime of the parameter space ($K < 1 $), 
we use the scaling relation eq.9 for the analysis of this figure.
the system
prefer the FM phase for the coupling $\lambda_1 $. As the value of $K$ increases the 
coupling $\lambda_2$ increases, i.e., the dqpII phase. This scaling
study is consistent with fig.4, as we increase the value of $K$, the coupling 
$\lambda_2$ increases and dqpII phase appears.\\
In the middle figure, the scaling results for $\lambda_1$ with $\mu$, 
we use the scaling relation eq.10 for the analysis of this figure.
It reveals from this
study for the strongly correlated regime that the order FM phase dominated over the 
dqpI phase. 
But
for the non-interacting and attractive regime the system 
prefers to be in the dqpI phase. 
This findings
of scaling results is consists with RG flow diagram study of fig. 1. \\
In right figure, we present the behaviour of the coupling $\lambda_2 $ with $\mu$,
we use the scaling relation eq.11 for the analysis of this figure. It reveals from
this scaling study that both of the couplings are proportional to each other for all
values of $K$, the green line actually is the superposition for the five different
values of K. This figure gives the evidence of co-existence phase for the
dqpI and dqpII. 
This results is consistent with the result of fig.6. \\ 
\section{Discussions:}
We have shown that 
in , correlated quantum Ising model, existence of    
an ordered ferromagnetic phase to disorder quantum paramagnetic phase transition 
occurs only for
strongly correlated regime otherwise the system is in the disorder quantum paramagnetic 
phase. 
We have also shown explicitly from the ordered ferromagnetic phase to disorder 
quantum paramagnetic quantum phase transition appeared for more correlated region
for the quantum Ising model with longer range coupling 
compare to the quantum Ising model.
We have also predicted two different kind of quantum phase transition one is
from the study of quantum Ising model and the other one is from the longer range
quantum Ising model. 
We have shown explicitly that the ferromagnetic coupling 
and transverse Ising field for the quantum Ising model are competing with each other, 
whereas for the longer range     
quantum Ising model, the transverse Ising field and longer range coupling are
not competing with each other. We have shown the existence of two different kinds
disorder quantum phase from two different roots. Our analysis of scaling relations
is also consistent with the results of quantum field theoretical RG results.   
This work provides a new perspective
not only in the statistical physics but also for the low dimensional quantum many body
physics.    
\section{ Method:  } 
{\bf (A). Derivation of Bosonized Hamiltonian } \\ 
We consider the Hamiltonian,
\begin{equation}
H = - \sum_{i}(\mu \sigma^{x}_{i}  + \lambda_{1} \sigma^{z}_{i} 
\sigma^{z}_{i+1}+ \lambda_{2} \sigma^{x}_{i}\sigma^{z}_{i-1}
\sigma^{z}_{i+1}),
\label{mh}
\end{equation} 
where $\lambda_{1}$ and $\lambda_{2}$ are the nearest and 
next nearest neighbor interactions. We write the Hamiltonian as,
\begin{equation}
H = - \sum_{i}(\mu S^{x}_{i}  + \lambda_{1} S^{z}_{i} S^{z}_{i+1}
+ \lambda_{2} S^{x}_{i}S^{z}_{i-1}S^{z}_{i+1}),
\label{mhs}
\end{equation} 
The field, $\phi$, corresponds to the spin fluctuations and
$\theta$ is the dual to the field $\phi$~\cite{gia,frad}.
These fields are related by the following relations
${\phi}_R = \theta - \phi $ and ${\phi}_L  = \theta + \phi $, are respectively
the right and left of the field. \\
The above two Hamiltonians are free from $K$.
Therefore, now our main task is to find the
analytical
expression for spin-1/2 operators in terms of in
terms of bosonized fields
$\phi$ and $\theta$ and that also show how $K$
appears in the quantum simulated model Hamiltonian.\\
We present spin operators interms of $\phi$, $\theta$ and $K$ $^{8,23,24}$,\\
We use the following transformation,
\begin{equation}
S_n^x = [\cos(2\sqrt{\pi K}\phi(x))+(-1)^n] \cos[\sqrt{\frac{\pi}{K}}\theta(x)] 
\end{equation}
\begin{equation}
 S_n^y = [\cos(2\sqrt{\pi K}\phi(x))+(-1)^n] \sin[\sqrt{\frac{\pi}{K}}\theta(x)] 
\end{equation}
\begin{equation}
S_n^z = (-1)^n \cos(2\sqrt{\pi K}\phi(x)) + \sqrt{\frac{\pi}{K}}\partial_x\phi(x) 
\end{equation}

\begin{multline}
S^{z}_{n} S^{z}_{n-1}= \left[ (-1)^n \cos(2\sqrt{\pi K}\phi(x)) 
+\sqrt{\frac{\pi}{K}}\partial_x\phi(x) \right]  \left[ (-1)^{n-1} 
\cos(2\sqrt{\pi K}\phi(x)) +\sqrt{\frac{\pi}{K}}\partial_x\phi(x) \right]\\
=(-1)^{2n-1}\cos^2(2\sqrt{\pi K}\phi(x)) +\frac{K}{\pi}(\partial_x\phi(x))^2
\end{multline}
\begin{multline}
S^{x}_{i}S^{z}_{i-1}S^{z}_{i+1}=S^{x}_{i}\left[ (-1)^n \cos(2\sqrt{\pi K}\phi(x)) 
+\sqrt{\frac{\pi}{K}}\partial_x\phi(x) \right] \\ 
\left[ (-1)^{n+1} \cos(2\sqrt{\pi K}\phi(x)) 
+\sqrt{\frac{\pi}{K}}\partial_x\phi(x) \right]\\
=[\cos^3(2\sqrt{\pi K}\phi(x))+(-1)^{n}\cos^2(2\sqrt{\pi K}\phi(x)) 
+\frac{K}{\pi}\cos^2(2\sqrt{\pi K}\phi(x))(\partial_x\phi(x))^2\\+(-1)^{n}
\sqrt{\frac{\pi}{K}}(\partial_x\phi(x))^2] \cos[\sqrt{\frac{\pi}{K}}\theta(x)]
\end{multline}
Neglecting oscillatory terms (i.e, $(-1)^n$) and higher order cosine 
terms (first term in the above equation), we have,
\begin{multline}
S^{x}_{i}S^{z}_{i-1}S^{z}_{i+1}=\cos(2\sqrt{\pi K}\phi(x))  
\cos[\sqrt{\frac{\pi}{K}}\theta(x)] (\sqrt{\frac{\pi}{K}}\partial_x\phi(x))^2
\end{multline}
We write final form of Bosonized Hamiltonian as,
\begin{multline}
H=H_0+\frac{\lambda_1}{2}\int \cos[4\sqrt{\pi}\phi(x)]dx - 
\mu \int 
\cos[2\sqrt{\pi}\phi(x)] \cos[\sqrt{\pi}\theta(x)] dx\\ -\frac{\lambda_2}{\pi} 
\int \cos[2\sqrt{\pi}\phi(x)] \cos[\sqrt{\pi}\theta(x)] \left( \partial_x \phi(x) 
\right)^2 dx 
\end{multline}
where $H_0 = \frac{v}{2} \int [(\partial_x \frac{1}{\sqrt{K}}
\theta)^2+ (\partial_x \sqrt{K} \phi)^2] dx $. In this derivation of RG calculation, we
use $K$ in $H_0$ instead of $V(\theta , \phi )$. \\
 
{\bf (B). Derivation of Renormalization Group Equation } \\ 
The action can be writte as,                                                            
\begin{dmath}
\left\langle S_{int}(\theta,\phi)\right\rangle =\frac{\lambda_1}{2}
\int \left\langle \cos[4\sqrt{\pi}\phi(r)]\right\rangle dr - \mu \int 
\left\langle \cos[2\sqrt{\pi}\phi(r)] \cos[\sqrt{\pi}\theta(r)] \right\rangle 
dr\\ -\frac{\lambda_2}{\pi} \int \left\langle \cos[2\sqrt{\pi}\phi(r)] 
\cos[\sqrt{\pi}\theta(r)] \left( \partial_r \phi(r) \right)^2\right\rangle  dr 
\end{dmath}
From the previous calculations we write the first order corrections,
\begin{equation}
\boxed{\frac{\lambda_1}{2}\int \left\langle \cos[4\sqrt{\pi}\phi(r)]
\right\rangle dr = \frac{\lambda_1}{2} b^{-4K} \int  \cos[4\sqrt{\pi}\phi_s(r)] dr }
\end{equation}
\begin{equation}
\boxed{\mu \int \left\langle \cos[2\sqrt{\pi}\phi(r)] \cos[\sqrt{\pi}\theta(r)] 
\right\rangle dr= \mu b^{-\left( K+\frac{1}{2K}\right) }\int dr 
\cos[2\sqrt{\pi}\phi_s(r)] \cos[\sqrt{\pi}\theta_s(r)] }
\end{equation}
\begin{multline}
\frac{\lambda_2}{\pi} \int \left\langle \cos[2\sqrt{\pi}\phi(r)] 
\cos[\sqrt{\pi}\theta(r)] \left( \partial_r \phi(r) \right)^2\right\rangle  
dr = (\frac{\lambda_2}{\pi}) \int \left\langle \cos[2\sqrt{\pi}\phi(r)] 
\cos[\sqrt{\pi}\theta(r)]\right\rangle  \left( \partial_r \phi_s(r) \right)^2 dr
\end{multline}
\begin{dmath}
=(\frac{\lambda_2}) {\pi}b^{-\left(K+\frac{1}{2K} \right)}  
\int dr \left(  \cos[2\sqrt{\pi}\phi_s(r)] \cos[\sqrt{\pi}\theta_s(r)]\right) .
\end{dmath}
Now we calculate the second order terms,
\begin{dmath}
-\frac{1}{2}\left( \left\langle S_{int}^2\right\rangle  - 
\left\langle S_{int}\right\rangle^2 \right) = -\frac{1}{2}
\int dr dr^{\prime} \left\lbrace \frac{\lambda_1^2}{4}[...]+\mu^2[...]
+\frac{\lambda_2^2}{\pi^2}
[...]-\frac{\lambda_1 \mu}{2}[...]-\frac{\lambda_1\lambda_2}{2\pi}[...]\\
-\frac{\mu\lambda_1}{2}
[...]+\frac{\mu\lambda_2}{\pi}[...]-\frac{\lambda_2\lambda_1}{2\pi}[...]
+\frac{\lambda_2\mu}{\pi}[...]\right\rbrace  
\end{dmath}
Following the previous calculations we write,
\begin{multline}
-\frac{\lambda_1^2}{8} \int dr dr^{\prime} \left\lbrace 
\left\langle \cos[4\sqrt{\pi}\phi(r)] \cos[4\sqrt{\pi}
\phi(r^{\prime})]\right\rangle - \left\langle 
\cos[4\sqrt{\pi}\phi(r)] \right\rangle \left\langle 
\cos[4\sqrt{\pi}\phi(r^{\prime})]\right\rangle  \right\rbrace \\
= \frac{\lambda_1^2}{8} \left( 1-b^{-8K} \right) \int dr \left( \partial_r \phi_s(r)\right)^2 
\end{multline}
\begin{multline}
-\frac{\mu^2}{2}\int dr dr^{\prime}  \left\langle 
\cos[2\sqrt{\pi}\phi(r)] \cos[\sqrt{\pi}\theta(r)] 
\cos[2\sqrt{\pi}\phi(r^{\prime})] \cos[\sqrt{\pi}
\theta(r^{\prime})] \right\rangle \\- \left\langle 
\cos[2\sqrt{\pi}\phi(r)] \cos[\sqrt{\pi}\theta(r)]\right\rangle \left\langle 
\cos[2\sqrt{\pi}\phi(r^{\prime})] \cos[\sqrt{\pi}\theta(r^{\prime})] \right\rangle \\
=  -\frac{\mu^2}{8} \left( b^{-4K}-b^{-\left( 2K-\frac{1}{2K}\right) }\right) 
\int dr \cos[4\sqrt{\pi}\phi_s(r)]
\end{multline}
\begin{multline}
\frac{\lambda_1\mu}{2}\int dr dr^{\prime}  \left\langle 
\cos[4\sqrt{\pi}\phi(r)] \cos[2\sqrt{\pi}\phi(r)] \cos[\sqrt{\pi}\theta(r)] 
\cos[4\sqrt{\pi}\phi(r^{\prime})] \cos[2\sqrt{\pi}\phi(r^{\prime})] 
\cos[\sqrt{\pi}\theta(r^{\prime})] \right\rangle \\- \left\langle 
\cos[4\sqrt{\pi}\phi(r)]\cos[2\sqrt{\pi}\phi(r)] 
\cos[\sqrt{\pi}\theta(r)]\right\rangle \left\langle 
\cos[4\sqrt{\pi}\phi(r^{\prime})] \cos[2\sqrt{\pi}\phi(r^{\prime})] 
\cos[\sqrt{\pi}\theta(r^{\prime})] \right\rangle \\
=  \frac{\lambda_1\mu}{4} \left( b^{-\left( K+\frac{1}{4K}\right) }
-b^{-\left( 5K-\frac{1}{4K}\right) }\right) \int dr \cos[2\sqrt{\pi}\phi_s(r)] 
\cos[\sqrt{\pi}\theta_s(r)]
\end{multline}
\begin{dmath}
-\frac{\lambda_2^2}{2\pi^2}\int dr dr^{\prime}  
\left\langle \cos[2\sqrt{\pi}\phi(r)] \cos[\sqrt{\pi}\theta(r)] 
\left( \partial_r\phi(r)\right)^2  \cos[2\sqrt{\pi}\phi(r^{\prime})]  
\cos[\sqrt{\pi}\theta(r^{\prime})] \left( \partial_r\phi(r)\right)^2\right\rangle \\
- \left\langle \cos[2\sqrt{\pi}\phi(r)] \cos[\sqrt{\pi}\theta(r)]
\left( \partial_r\phi(r)\right)^2\right\rangle \left\langle 
\cos[2\sqrt{\pi}\phi(r^{\prime})]  \cos[\sqrt{\pi}\theta(r^{\prime})] 
\left( \partial_r\phi(r^{\prime})\right)^2\right\rangle \\
=  -\frac{\lambda_2^2}{2\pi^2} \int dr dr^{\prime} \left\lbrace 
\left( \left\langle \cos[r] \cos[r^{\prime}] \right\rangle 
- \left\langle \cos(r)\right\rangle \left\langle \cos(r^{\prime})\right)\right\rangle  
\left( \partial_r\phi_s(r)\right)^4 
+ \left( \left\langle \cos(r)\cos(r^{\prime}) 
\left( \partial_r^{\prime}\phi_f(r^{\prime})\right)^2\right\rangle 
- \left\langle \cos(r)\right\rangle \left\langle \cos(r^{\prime})
\left( \partial_r\phi_f(r^{\prime})\right)^2\right\rangle \right) 
\left( \partial_r\phi_s(r)\right)^2 
+ 2\left( \left\langle \cos(r)\cos(r^{\prime}) \left( \partial_r^{\prime}\phi_f(r^{\prime})\right)
\right\rangle - \left\langle \cos(r)\right\rangle \left\langle \cos(r^{\prime})\left( \partial_r\phi_f(r^{\prime})\right)\right\rangle \right) \left( \partial_r\phi_s(r)\right) \left( \partial_r^{\prime}\phi_s(r^{\prime})\right)
+\left( \left\langle \cos(r)\cos(r^{\prime}) \left( \partial_r\phi_f(r)\right)^2\right\rangle 
- \left\langle \cos(r) \left( \partial_r\phi_f(r)\right)\right\rangle 
\left\langle \cos(r^{\prime})\right\rangle \right) \left( \partial_r^{\prime}\phi_s(r^{\prime})\right)^2
+\left( \left\langle \cos(r)\cos(r^{\prime}) 
\left( \partial_r\phi_f(r)\right)^2 \left( \partial_r^{\prime}\phi_f(r^{\prime})\right)^2
\right\rangle - \left\langle \cos(r) \left( \partial_r\phi_f(r)\right)^2
\right\rangle \left\langle \cos(r^{\prime}) \left( \partial_r^{\prime}
\phi_s(r^{\prime})\right)^2\right) \right\rangle
+2 \left( \left\langle \cos(r)\cos(r^{\prime}) \left( \partial_r\phi_f(r)\right)^2 
\left( \partial_r^{\prime}\phi_f(r^{\prime})\right)\right\rangle 
- \left\langle \cos(r) \left( \partial_r\phi_f(r)\right)^2\right\rangle 
\left\langle \cos(r^{\prime}) \left( \partial_r^{\prime}\phi_s(r^{\prime})\right)\right)
 \right\rangle \left( \partial_r^{\prime}\phi_s(r^{\prime})\right) 
+2 \left( \left\langle \cos(r)\cos(r^{\prime}) \left( \partial_r\phi_f(r)\right) \right\rangle 
- \left\langle \cos(r) \left( \partial_r\phi_f(r)\right)\right\rangle \left\langle \cos(r^{\prime})
\right\rangle \right)\left( \partial_r\phi_s(r)\right)\left( \partial_r^{\prime}
\phi_s(r^{\prime})\right)^2 
+2 \left( \left\langle \cos(r)\cos(r^{\prime}) \left( \partial_r\phi_f(r)\right) \left( \partial_r^{\prime}\phi_f(r^{\prime})\right)^2\right\rangle - \left\langle \cos(r) \left( \partial_r\phi_f(r)\right)^2\right\rangle \left\langle \cos(r^{\prime}) \left( \partial_r^{\prime}\phi_s(r^{\prime})\right)^2\right) \right\rangle 
\left( \partial_r\phi_s(r)\right) 
+4 \left( \left\langle \cos(r)\cos(r^{\prime}) \left( \partial_r\phi_f(r)\right) \left( \partial_r^{\prime}\phi_f(r^{\prime})\right)\right\rangle - \left\langle \cos(r)
 \left( \partial_r\phi_f(r)\right)\right\rangle \left\langle \cos(r^{\prime}) \left( \partial_r^{\prime}\phi_s(r^{\prime})\right)\right) \right\rangle 
 \left( \partial_r\phi_s(r)\right) \left( \partial_r^{\prime}\phi_s(r^{\prime})\right)
\right\rbrace \\
= 0 
\end{dmath}
Thus all the correlation functions vanish and $\lambda_2^2 $ 
term goes to zero.
Now we calculate $\lambda_1\lambda_2$ term.
\begin{multline}
\frac{\lambda_1\lambda_2}{4\pi}\int dr dr^{\prime}  \left\langle 
\cos[4\sqrt{\pi}\phi(r)] \cos[2\sqrt{\pi}\phi(r^{\prime})]
\cos[\sqrt{\pi}\theta(r^{\prime})] \left( \partial_r^{\prime}
\phi(r^{\prime})\right)^2 \right\rangle \\- \left\langle 
\cos[4\sqrt{\pi}\phi(r)] \right\rangle \left\langle  
\cos[2\sqrt{\pi}\phi(r^{\prime})] \cos[\sqrt{\pi}\theta(r^{\prime})]
\left( \partial_r^{\prime}\phi(r^{\prime})\right)^2\right\rangle 
\end{multline}
Here correlation function,
\begin{multline}
\left\langle \cos[4\sqrt{\pi}\phi(r)] \cos[2\sqrt{\pi}\phi(r^{\prime})]
\cos[\sqrt{\pi}\theta(r^{\prime})] \left( \partial_r^{\prime}
\phi(r^{\prime})\right)^2 \right\rangle \\= \left\langle 
\cos[4\sqrt{\pi}\phi(r)] \cos[2\sqrt{\pi}\phi(r^{\prime})]
\cos[\sqrt{\pi}\theta(r^{\prime})]  \right\rangle 
\left( \partial_r^{\prime}\phi_s(r^{\prime})\right)^2
\end{multline}
similarly we have $\left\langle \cos[4\sqrt{\pi}\phi(r)] 
\right\rangle \left\langle \cos[2\sqrt{\pi}\phi(r^{\prime})]
\cos[\sqrt{\pi}\theta(r^{\prime})]  \right\rangle 
\left( \partial_r^{\prime}\phi_s(r^{\prime})\right)^2$. 
Thus we have,
\begin{multline}
\frac{\lambda_1\lambda_2}{4\pi}\int dr dr^{\prime} \\ 
\left\lbrace \left\langle \cos[4\sqrt{\pi}\phi(r)] 
\cos[2\sqrt{\pi}\phi(r^{\prime})]\cos[\sqrt{\pi}\theta(r^{\prime})]  
\right\rangle - \left\langle \cos[4\sqrt{\pi}\phi(r)] 
\right\rangle \left\langle \cos[2\sqrt{\pi}\phi(r^{\prime})]
\cos[\sqrt{\pi}\theta(r^{\prime})]  \right\rangle\right\rbrace\\ 
\left( \partial_r^{\prime}\phi_s(r^{\prime})\right)^2 
= \frac{\lambda_1\lambda_2}{4\pi} \left( b^{-\left( K+\frac{1}{4K}\right) }
-b^{-\left( 5K-\frac{1}{4K}\right) }\right) \int dr 
\cos[2\sqrt{\pi}\phi(r)]\cos[\sqrt{\pi}\theta(r)] \left( \partial_r^{\prime}
\phi_s(r^{\prime})\right)^2
\end{multline}
Finally we obtain our RG equations for the Hamiltonian, $ H$ (eq.20), 
using the above all equations:\\
\begin{equation}
\frac{d\lambda_1}{dl}=\left(2-4K \right)\lambda_1
+\frac{\mu^2}{8}\left(2K-\frac{1}{2K} \right) \nonumber\\ 
\end{equation}
\begin{equation}
\frac{d\mu}{dl}=\left(2-K-\frac{1}{4K} \right)\mu 
+\lambda_1\mu K  \nonumber\\ 
\end{equation}
\begin{equation}
\frac{d\lambda_2}{dl}=\left(2-K-\frac{1}{4K} \right)\lambda_2  
+\frac{\lambda_1\lambda_2 K}{\pi} \nonumber\\ 
\end{equation}
\begin{equation}
\frac{dK}{dl}=-\lambda_1^2 K^2
\end{equation}

Similarly one can find the RG equation only for the coupling $\lambda_1 $. 
\begin{equation}
\frac{d\lambda_1}{dl}=\left(2-4K \right)\lambda_1
+\frac{\mu^2}{8}\left(2K-\frac{1}{2K} \right) \nonumber\\ 
\end{equation}
\begin{equation}
\frac{d\mu}{dl}=\left(2-K-\frac{1}{4K} \right)\mu 
+\lambda_1\mu K  \nonumber\\ 
\end{equation}
\begin{equation}
\frac{dK}{dl}=-\lambda_1^2 K^2
\end{equation}

Similarly one can find the RG equation in absence of $\mu (=0)$.\\
\begin{equation}
\frac{d\lambda_1}{dl}=\left(2-4K \right)\lambda_1
\nonumber\\ 
\end{equation}
\begin{equation}
\frac{d\lambda_2}{dl}=\left(2-K-\frac{1}{4K} \right)\lambda_2  
+\frac{\lambda_1\lambda_2 K}{\pi} \nonumber\\ 
\end{equation}
\begin{equation}
\frac{dK}{dl}=-\lambda_1^2 K^2
\end{equation}

\textbf{Acknowledgments}\\
The authors acknowledge 
Prof. Chandan Dasgupta, Prof. Prabir Mukherji
for 
critically reviewing the manuscript.\\ 
The author would like to acknowledge AMEF, 
DST (EMR/2017/000898) and RRI library for 
books and journals.\\
\textbf{Author contributions statement:}\\
S.S. identified the problem and also write the manuscript, R.R.K. do the calculations of this problem 
under the guidence of S.S. 
All authors analysed the results and reviewed the manuscript. \\
{\bf Additional Informations: } \\
Competing interests: The author declare no competing interests.

\end{document}